# Femtosecond electron and x-ray source based on laser wakefield accelerator

Dmitri A. Oulianov [a], Yuelin Li [b], Robert A. Crowell [a], Ilya A. Shkrob [a], David J. Gosztola [a],
Oleg J. Korovyanko [a], Roberto C. Rey-de-Castro [a]
[a] Chemistry Division, Argonne National Laboratory, Argonne, IL 60439
[b] Accelerator Systems Division, Argonne National Laboratory, Argonne, IL 60439

**ABSTRACT**

A terawatt tabletop laser wakefield acceleration source of relativistic electrons has been developed in our Terawatt Ultrafast High Field Facility (TUHFF). The preliminary results for ultrafast radiolysis of liquid water using this femtosecond electron source are presented. A TUHFF based femtosecond x-ray source is proposed. Thomson scattering of the accelerated electrons off a counterpropagating terawatt laser beam will be used to generate keV x-ray photons. The expected parameters of this x-ray source have been estimated. The short pulse duration, high flux, and good collimation of the resulting x-ray beam would be conducive for ultrafast time-resolved x-ray absorption studies of short-lived transient species in gases, liquids, and solids. It is argued that the solvation dynamics of Br atoms generated in photoinduced electron detachment from aqueous bromide would make a convenient choice for the first pump-probe experiment using this x-ray source.

## 1. INTRODUCTION

Short-pulse sources of x-ray radiation and relativistic electrons would undoubtedly have a large impact on the studies of material properties. Low energy (1-100 keV) femtosecond electron pulses are perfect for ultrafast electron diffraction[1], electron pulses with 0.5-50 MeV could be used for pulse radiolysis[2], while electron pulses with higher energies (from 100 MeV to GeV) can be used for high energy physics research[3-5]. The application of femtosecond x-ray pulses[6-12] for ultrafast studies in physics and chemistry allows one to observe atomic motion and structural changes in the condensed phase on the time scale of a single vibrational period. The use of x-rays as a probe is particularly attractive because the x-ray photons interact with core electrons in the constituent atoms and thereby provide direct information about electronic and atomic structure. In addition, by tuning the x-ray photon energy to a resonance with a specific atom, it is possible to selectively observe certain types of atoms in complex materials. Several nanosecond and picosecond x-ray techniques based on synchrotron radiation and laser sources have been developed during the past decade[13]. Most of these techniques use an ultrashort laser pulse to initiate a photoreaction followed by a delayed pulse of x-rays that probes the subsequent structural changes. In metal and semiconductor crystals, nonthermal melting and laser-induced heat and strain propagation have been studied by means of time-resolved x-ray diffraction on the nanosecond and picosecond time scale[14,15]. Excited-state structures of photoactive proteins have been determined using nanosecond x-ray crystallography.[16,17] Picosecond x-ray powder diffraction was used to observe short-lived electronically excited states in organic solids[18]. Photodissociation of molecular iodine in a liquid solution was studied using picosecond x-ray scattering[19]. Picosecond and nanosecond Extended X-ray Absorption Fine Structure (EXAFS) spectroscopy has been developed to provide information about short-range structures and coordination of atoms of molecular excited states in disordered media, such as liquids and amorphous solids[20-24]. The next step is to improve this time resolution from picoseconds to a few tens of femtoseconds.

There are several different ways to generate femtosecond x-ray pulses[6-12]. One of these methods is Thomson scattering of an intense laser pulse from an electron beam. The generation of 300 fs FWHM hard x-ray pulses using 90° Thomson scattering between femtosecond laser pulses and tightly focused nanosecond synchrotron electron bunches have been demonstrated[7]. The relatively low photon flux of ~$10^6$ photons/s obtained in these experiments could be significantly increased using more powerful laser sources, optimizing the experimental design and using high quality femtosecond electron bunches. Laser wakefield accelerators can produce such intense bunches. It was shown theoretically[10] and experimentally[9] that such accelerators can produce highly directional x-ray beam with a flux of ~$10^{11}$ photons/s and <100 fs pulse duration. The x-ray energy spectrum depends on the energy distribution of the electrons, which is different in different acceleration regimes[3-5,25]. In the self-modulated laser wakefield (SMLWF) regime the calculated[10] x-ray spectrum is very broad; with a maximum at 2-3 keV and a tail extending to 20 keV. Phuoc et al.[9] used forced laser wakefield (FLWF) regime of acceleration and generated x-rays with a spectrum that exhibited a maximum at 100 eV and a tail extending to 2 keV. This spectrum was attributed to Thomson scattering of co-propagating laser and electron beams. Their detector, however, was not sensitive to x-rays with higher energies. Both of these approaches have produced high flux, low divergence x-ray beams with femtosecond pulse duration and broad energy spectrum that are optimum for ultrafast x-ray absorption spectroscopy experiments.

In this paper we describe a source of femtosecond keV x-ray pulses that is based on Thomson scattering using terawatt laser wakefield accelerator developed at our Terawatt Ultrafast High Field Facility (TUHFF). TUHFF houses a 20 terawatt laser system used to generate subpicosecond pulses of relativistic electrons with the total charge of ~3 nC per pulse. We present calculated parameters of this x-ray source based on the achieved electron beam characteristics.

We also discuss pulse radiolysis experiments in which we use the electron beam directly as a pump source for chemical studies. The interaction between ionizing radiation and the molecules in the sample initiates and drives the chemical processes. A fundamental understanding of these interactions is relevant to such diverse areas as environmental chemistry, astrochemistry, medical radiotherapy, stability of nuclear waste repositories and to a wide variety of technologically important processes such as plasma enhanced vapor deposition. The outcomes of most radiation induced chemical reactions are determined by ultrafast processes such as energy transfer, thermalization, and solvation.

Due to the lack of a suitable short-pulse source of ionizing radiation, experimental studies on the primary events of radiation induced chemical reactions are nonexistent. While ultrafast laser studies provide many new insights, the results clearly demonstrate that lasers are not capable of fully reproducing the chemistry that is specific to ionizing radiation. This is primarily a result of different mechanisms of energy deposition in the condensed phase. Laser photo-ionization produces isolated ionization events of a specific energy. In pulse radiolysis tremendous amounts of energy are deposited in small volumes; the resulting clusters of ionized and excited molecules are called spurs (e.g., 1-5 nm diameter in water). Combined with the lack of selection rules and the inhomogeneous nature of the energy deposition within the spur, an assortment of ions, radicals and electrons are produced in a variety of energetic states and in close proximity of each other. Detailed knowledge of the ultrafast primary events within the spur is essential in order to produce a complete understanding of radiation chemistry.

To address this issue the electron beam generated by our laser wakefield accelerator was used to ionize liquid water in our setup. The preliminary liquid water radiolysis data are described as the first demonstration experiment.

## 2. TERAWATT LASER SYSTEM

Our 20 TW laser system is described only briefly here (see ref. 2 for more detail). The laser system consists of a mode-locked Ti:Sapphire oscillator (KM Labs Basic), a stretcher, three multi-pass Ti:Sapphire amplifiers and a compressor. The oscillator is pumped with 4.75W output from a diode pumped frequency doubled $Nd:YVO_4$ laser (Spectra Physics Millennia V). The output of the oscillator exhibits a broad spectrum (FWHM ~50nm) that is centered at 775nm. The oscillator produces 5nJ pulses at a repetition rate of 100MHz. The pulses are then passed through a 1:1 imaging single grating stretcher to obtain a pulsewidth of ~300ps. The pulse train is then passed through a 10Hz pulse picker. The latter consists of a Pockel cell polarizer (Medox). The first amplifier has a multi-pass ring design. It houses a 1cm thick Brewster angle amplifier rod (Crystal Systems) that is pumped using 25mJ (532nm) light that is split off of the output of a frequency-doubled Nd:YAG (Spectra Physics PRO-270). After six passes through the amplifier the IR pulse is passed through a second pulse picker that eliminates any amplified spontaneous emission background. The pulse energy at this point is 2.5mJ; the beam diameter is 0.8 cm ($1/e^2$). The second amplifier rod (Crystal Systems) is an anti-reflection coated 2cm cylinder. Two frequency-doubled beams from Nd:YAG lasers (Spectra Physics LAB-190 and Spectra Physics PRO-270) are relay imaged onto each face of the amplifier rod. The total pump energy is 850mJ. This amplifier has a four pass bow tie configuration. A -1.5m lens is placed after the first pass to compensate for the thermal lensing that occurs in the amplifier rod. The output of the second amplifier is passed through a spatial filter that expands the beam diameter to 1.2cm ($1/e^2$). At this point, the pulse energy is 330mJ. A beamsplitter is used to divert ~60mJ of this beam which is passed through a second compressor and used as a probe. The third amplifier rod (Crystal Systems) is an anti-reflection coated 3cm cylinder. The third amplifier is pumped on each face by a relay imaged frequency-doubled Nd:YAG laser (Spectra Physics PRO-350). The total pump energy is 2.4J. To eliminate thermal lensing in the third amplifier the rod is mounted in a liquid nitrogen cryostat (Janis Cyrogenics). Two passes through the amplifier boost the pulse energy to 1.1J. After the first pass slight aberrations in the beam profile are cancelled out by rotation of the beam profile by 90º before the second pass. After the third amplifier, the beam is expanded to 5cm and directed into a vacuum chamber that houses a two-grating pulse compressor. The output pulse has a spectrum that is centered at 800 nm and has a width of 34 nm FWHM. The compressed pulse is 600 mJ, 35fs FWHM. The repetition rate of the setup is 1-10 Hz. Uultrashort electron pulses are generated by focusing the resulting laser beam onto the front surface of a 1.2 mm supersonic He jet using a 50 cm focal length off-axis gold-plated parabolic mirror. The jet characteristics are crucial for obtaining the high electron output. We have tested different types of nozzle designs and selected the one which produces the highest electron charge. This nozzle has a similar design to the one used by Umstadter group.[26] For electron charge measurements a home-built 4.5 cm diameter Faraday cup was used; this cup was placed ca. 2 cm after the jet. In order to block the laser light and low energy electrons the Faraday cup is screened using a 80 μm thick Al window. The total electron charge is 2-3 nC per pulse with a pulse-to-pulse variation of 15-30%. This charge is distributed over a wide cone of almost 20º (Fig. 1). Within this cone there is a narrower 6º cone of the electrons that have somewhat higher energies (see below). The electrons generated in the plasma undergo multiple elastic scattering events as they emerge from the jet. This scattering is stronger for low energy electrons and it results in a steep angular distribution of electron energies.

The electron beam spectrum was determined using a spectrometer in which two permanent magnets were used to bend the electron beam. The electrons were imaged on the luminescent screen and the emitted light was observed using a camera. Copper apertures were used to dissect the electron beam. The magnetic field of the spectrometer was mapped and relativistic equations solved numerically to reconstruct the energy spectrum. Figs. 2a and 2b show the spectra obtained for electrons scattered into 2º and 14º cones, respectively. In the latter figure, the most probable electron energy is ca. 4 MeV. There is also a sloping "tail" extending to at least 25 MeV. In the 2º cone, the most probable energy is increased to 7-8 MeV and the fraction of low-energy electrons is greatly reduced. Very similar energy distributions can be obtained theoretically assuming Maxwell distribution for the electron energies and small angle multiple elastic scattering in the jet. To characterize the average energy of the electrons across the whole beam the latter was passed through a 1 cm thick stack of aluminum plates and the distribution of the dose was determined using a radiochromic film inserted between the plates. The dose distribution was consistent with a Maxwell distribution of electron energies with the average energy of 2.3±0.3 MeV.

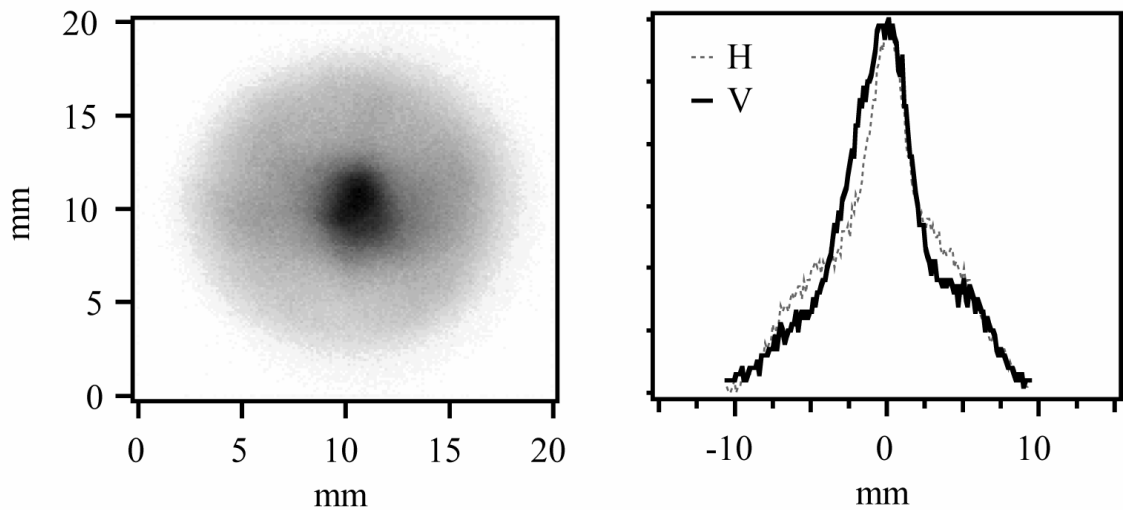

Figure 1: Transverse profile of the electron beam measured at 27 mm after the jet. The beam profile is almost identical in both the vertical (V) and horizontal (H) planes.

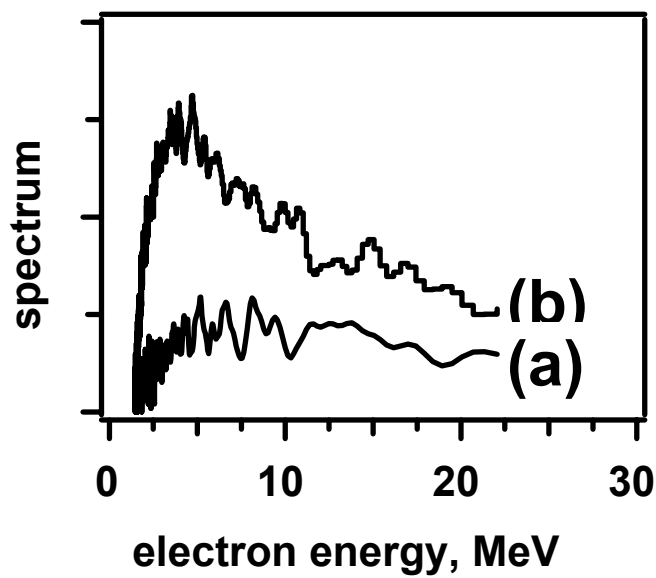

Figure 2: The energy spectrum of the electron beam measured for (a) 2° and (b) 14° cones.

The narrow cone contains the electrons that would be optimum for doing ultrafast experiments, but the fraction of these high-energy electrons in the beam is too low for 90° crossed-beam spectroscopy described below. The scattering in the jet can be reduced by optimizing the jet parameters (e.g., by reducing the gas density and increasing the gas velocity), which will be done in the future.

## 2. ULTRAFAST PULSE RADIOLYSIS

In the pulse radiolysis experiment (Fig. 3), a 80 μm thick Al foil is placed at 1.7 cm after the jet to block the laser light and transmit MeV electrons. The transmitted electron beam is used to radiolize the sample in a pump-probe experiment. It passes through a 1 cm path Suprasil cell filled with liquid water; the front window of this cell is 1 cm away from the Al shield. The electron beam is crossed at 90° inside the cell with two collinear probe beams: 800 nm 35 fs beam and a 670 nm cw laser beam. The first probe beam is derived from the 2nd amplifier. After passing through a variable length compressor, and a motorized delay stage this probe beam is split into two beams by a 50% beam splitter. The transmitted beam is used as a reference, while the reflected beam is focused into a sample cell and used as a probe "signal" beam. The second probe beam, derived from a 670 nm cw diode laser, is used to compensate for pulse-to-pulse variations in the absorbed dose. It propagates along the same optical path as the 800 nm probe beam. Both of these probe beams are used to measure transient absorption of solvated electrons generated by the electron pulse. However, 800 nm beam is used to obtain the transient absorption on the picosecond time scale in a stroboscopic fashion, whereas the 670 nm cw beam registers the same absorption on the nanosecond to microsecond time scales using a 30MHz photodiode and a fast transient digitizer. The solvated electron in water is long lived and both absorbance signals scale linearly with the dose. If both of the beams probe the same region of the sample, the two TA signals track each other. This proportionality can be used to compensate the variation in the dose deposition. The time resolution of our setup is determined by the optical path of the 800 nm probe beam across the radiolized region of the sample.

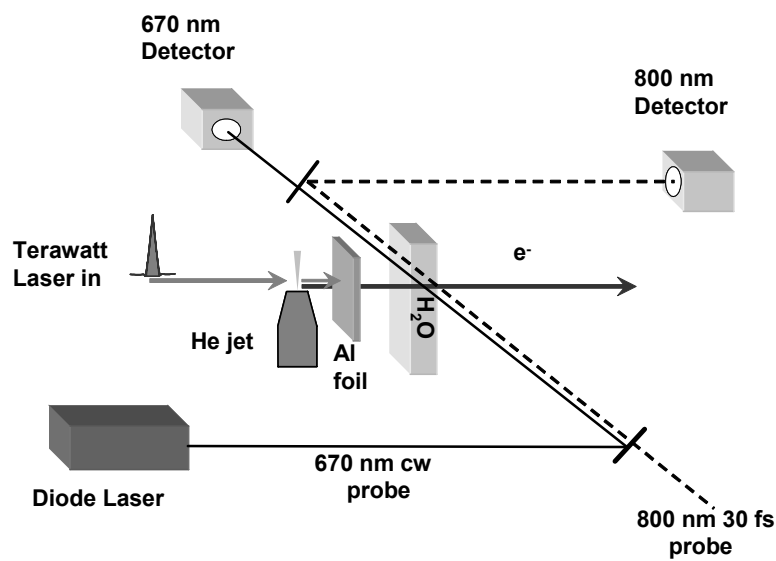

Figure 3: Ultrafast pulse radiolysis setup

The electron beam diameter at the sample is ca. 0.55 cm FWHM (see Fig. 1), which corresponds to ~30 ps rise time. This time resolution can be improved by using an aperture which transmits only the central part of the electron beam.

In our first demonstration experiment transient absorption of solvated electron in radiolized water was observed. The absorption spectrum and decay kinetics of this species are well known from the previous work[27]. Fig. 4 shows the decay kinetics of the solvated electron on the microsecond time scale following a 75pC electron pulse from TUHFF. This decay is due to the reaction of the solvated electron with electron-acceptor impurities in water, mainly traces of oxygen.

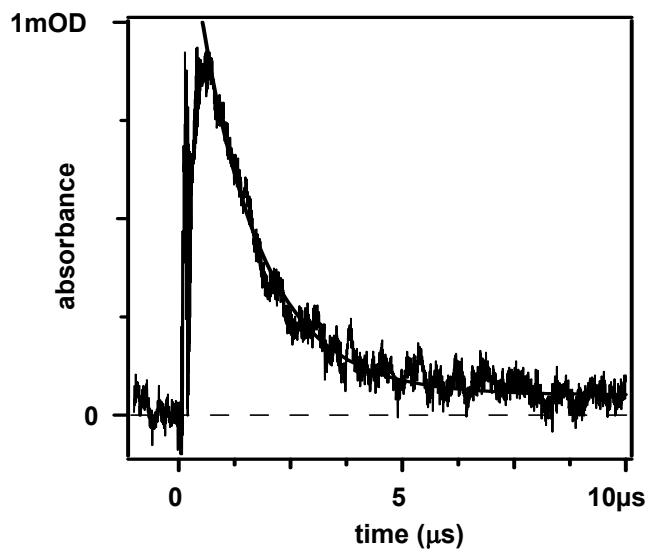

Figure 4: Kinetics of the solvated electron in water measured by a 670 nm CW probe beam.

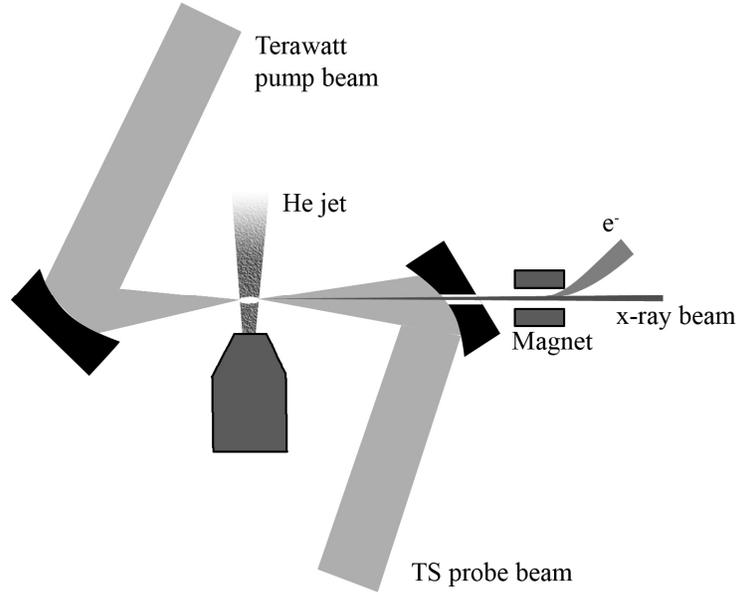

Figure 5: Thomson scattering x-ray source

The experiments on ultrafast radiolysis with 800 nm 35fs probe are currently in progress. A transient absorption signal of 18±1 mOD from the solvated electron was observed at the delay time of 700 ps following a 2 nC electron pulse from TUHFF.

## 3. THOMSON SCATTERING X-RAY SOURCE

TUHFF can also be used to generate x-rays via Thomson scattering. The setup is shown in Fig. 5. In this setup the probe beam (30 mJ/pulse, 30 fs) is focused onto the back surface of the jet using the second parabolic mirror. This mirror has a small hole at the center for the electrons and the x-rays to pass through. A bending magnet placed behind the parabolic mirror removes the electrons out of the beam. For a head-on collision, the system has an azimuthal symmetry. The scattering cross section after integration over the azimuthal angle is

$$\frac{d\Sigma}{d\theta} = \frac{\pi r_e^2}{\gamma^2} \frac{1}{(1-\beta\cos\theta)^2}\left[1+\left(\frac{\cos\theta-\beta}{1-\beta\cos\theta}\right)^2\right]\sin\theta. \qquad (1)$$

Here $\theta$ is the angle between the electron propagation direction and the scattered photons, $\gamma=(1-\beta^2)^{-1/2}$ is the relativistic factor of the electron beam, and $r_e=e^2/mc^2=2.82\times10^{-13}$ cm is the classical electron radius. The energy relation between the scattered and the incident photons is

$$\varepsilon_x = \varepsilon_i \frac{1+\beta}{1-\beta\cos\theta}, \qquad (2)$$

for incident photon energy $\varepsilon_i$ that is much smaller than the electron energy. The integration is assuming that the distribution of the electron energies is given by

$$f(\gamma) = \frac{1}{\gamma_o} e^{-\frac{\gamma}{\gamma_o}}.$$ (3)

Here $\gamma_0$ is determined from the experiment, and the bandwidth $\Delta\gamma/\gamma=(\Delta\omega/\omega)^{1/2}$. The total photon flux is obtained by integrating over time and over all rays in the electron and laser beams to obtain

$$F_T = \frac{N_e N_{ph}}{2\sqrt{2}\pi r_0^2},$$ (4)

where $N_e$ and $N_{ph}$ are the total number of electrons and photons, $r_0$ is the transverse beam size of the both. Hence the photon flux in a cone angle of $\Theta$ is given by

$$F(\varepsilon_x) = F_T \frac{\Delta\gamma}{\gamma} \int_0^\Theta \frac{d\Sigma}{d\theta} d\theta.$$ (5)

Here the integration is carried out using Eq. (2) for a fixed $\varepsilon_x$. The debunch due to the energy spread is neglected as it would not affect the peak brightness at the high energy end of the electron spectrum. Band width of the laser is ignored due to the much larger energy spread of the electron energy. Divergence of the electron beam is not taken into account in the calculation.

The results of this calculation are shown in Figs. 6a and 6b. The parameters used in the calculation are given in Table 1.

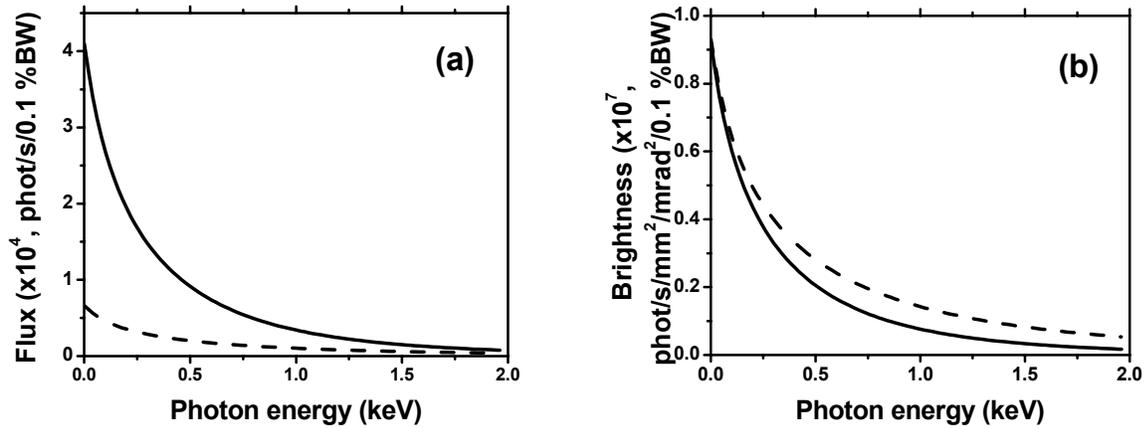

Figure 6: Average x-ray flux (a) and average brightness (b) into 50 (solid line) and 20 (dash line) mrad collection angles for the electron beam with 2MeV temperature.

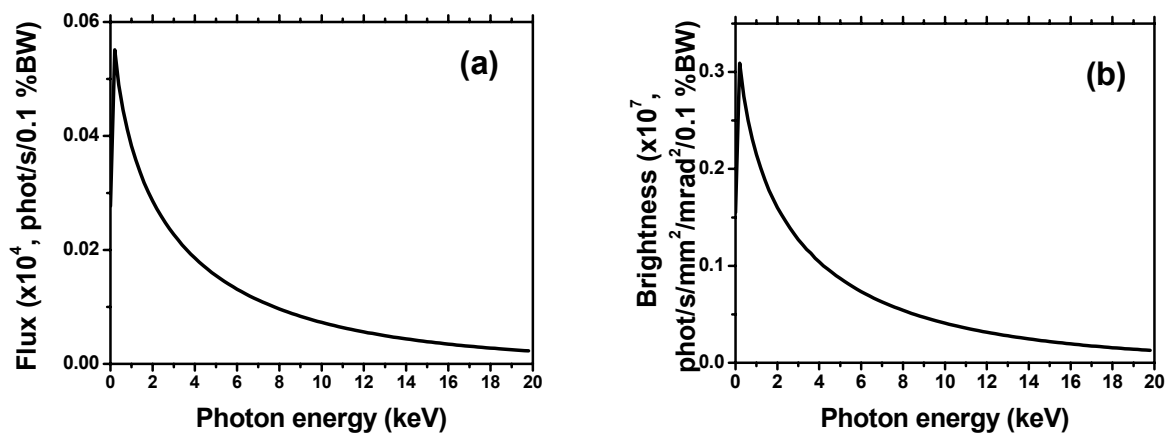

Figure 7: Average x-ray flux (a) and average brightness (b) into 10 mrad collection angle for the electron beam with 20MeV temperature.

Table 1: Parameters for Thomson scattering, assuming a 50 mrad cone angle

| | |
|---|---|
| Laser photon energy $\varepsilon_i$ | 1.55 eV |
| Laser pulse photon number $N_{ph}$ (energy) | $1.2 \times 10^{17}$ (30 mJ) |
| Laser pulse duration | 30 fs |
| Eleectron beam energy $\gamma$ | 6 |
| Number of electrons $N_e$ (charge) | $1.25 \times 10^{10}$ (2 nC) |
| Bunch length | 1 ps |
| Beam size $r_0$ | 10 $\mu$m |
| Repetition rate | 10 Hz |
| Average X-ray flux | $10^4$ ph/s /0.1%BW |
| Average X-ray brightness | $10^7$ ph/s/mm$^2$/mrad$^2$/0.1%BW |
| Peak brightness | $10^{19}$ ph/s/mm$^2$/mrad$^2$/0.1%BW |
| Total photon number | $6 \times 10^8$ ph/s |

According to this calculation, the spectrum of the Thomson scattering x-ray source that uses the electron beam currently generated by TUHFF will extend to 1-2 keV (which is consistent with the experimental data presented in ref. 9). Such an x-ray source might be used in surface or gas phase studies, but it is not useful for studies of bulk samples. In order to obtain higher x-ray energies it is necessary to increase the plasma temperature, which may be possible after the jet design optimization[3-5,25]. Fig. 7 shows the results for the electron beam with the mean temperature of 20 MeV (all other parameters being the same). This temperature was reported by Malka et al.[25] who used a similar setup.

Ultrafast dynamics of halogen atoms in charge-transfer-to-solvent (CTTS) photoreactions of aqueous halide anions may provide a convenient test bed for our TUHFF based femtosecond x-ray source. This CTTS photoreaction provides the simplest example of medium-assisted electron transfer. As such, it has been extensively studied both theroretically and experimentally. Despite of these many studies, little is yet known about the dynamics of halogen atom solvation in the course of this important photoreaction. The halogen atoms are formed when the electron is detached from the parent anion. By using pump-probe K- or L-edge EXAFS spectroscopy it will be possible to observe directly the structural changes around the halogen atom/anion that occur immediately after the electron transfer on the femtosecond time scale.

We believe that the CTTS reaction involving Br- would be suitable for a pump-probe K-band EXAFS study using our femtosecond x-ray source, because: 1) the solvation of Br atom is expected to occur on sub-picosecond time scale. 2) The local structure of the solvent around the halogen is expected to change considerably after the electron detachment. For example, in I⁻(H$_2$O)n clusters, a change of ~0.2 Å in the I-O distance was predicted theoretically[28]. The oxidation dramatically changes the x-ray absorption spectrum of the halide anions. 3) The intermediate states and electron detachment kinetics of the CTTS reaction are well studied by optical spectroscopy methods. 4) There are detailed theoretical models of the CTTS photoreaction that can be readily tested against the experimental data.

## ACKNOWLEDGMENT

This work was supported by the Office of Basic Energy Sciences, Division of Chemical Sciences, U. S. Department of Energy under Contract number W-31-109-Eng-38. We would like to thank C.D. Jonah for valuable discussions.